\documentstyle[12pt]{article}

\newcommand{\sect}[1]{\setcounter{equation}{0}\section{#1}}

\newcommand{\EQ}{\begin{equation}}
\newcommand{\EN}{\end{equation}}
\newcommand{\bea}{\begin{eqnarray}}
\newcommand{\ena}{\end{eqnarray}}

\renewcommand{\d}{\delta}

\newcommand{\D}{\Delta}

\newcommand{\p}{\pi}

\renewcommand{\t}{\tau}

\def\IP{\relax{\rm I\kern-.18em P}}

\newcommand{\shalf}{\frac{1}{2}}

\newcommand{\NP}[1]{Nucl.\ Phys.\ {\bf #1}}
\newcommand{\PL}[1]{Phys.\ Lett.\ {\bf #1}}

\def\bfone{\relax{\rm 1\kern-.35em 1}}
\def\bfzero{\relax{\rm I\kern-.18em 0}}
\def\inbar{\vrule height1.5ex width.4pt depth0pt}
\def\IC{\relax\,\hbox{$\inbar\kern-.3em{\rm C}$}}
\def\ID{\relax{\rm I\kern-.18em D}}
\def\IF{\relax{\rm I\kern-.18em F}}
\def\IK{\relax{\rm I\kern-.18em K}}
\def\IH{\relax{\rm I\kern-.18em H}}
\def\II{\relax{\rm I\kern-.17em I}}
\def\IN{\relax{\rm I\kern-.18em N}}
\def\IQ{\relax\,\hbox{$\inbar\kern-.3em{\rm Q}$}}
\def\IR{\relax{\rm I\kern-.18em R}}
\def\IG{\relax\,\hbox{$\inbar\kern-.3em{\rm G}$}}
\font\cmss=cmss10 \font\cmsss=cmss10 at 7pt
\def\ZZ{\relax\ifmmode\mathchoice
{\hbox{\cmss Z\kern-.4em Z}}{\hbox{\cmss Z\kern-.4em Z}}
{\lower.9pt\hbox{\cmsss Z\kern-.4em Z}}
{\lower1.2pt\hbox{\cmsss Z\kern-.4em Z}}\else{\cmss Z\kern-.4em
Z}\fi}

\begin{document}

\begin{titlepage}
\begin{flushright} IFUM-595-FT
\end{flushright}
\vfill
\begin{center}
{\LARGE\bf Non--transversal colliding singularities in F--theory
\footnote{To appear in the proceedings of {\em Quantum aspects of 
gauge theories, supersymmetry and unification}, Neuchatel 
University, 18--23 September 1997}}\\
\vskip 27.mm  \large
{\bf   Silvia Penati, Alberto Santambrogio and 
Daniela Zanon } \\
\vfill
{\small
Dipartimento di Fisica dell'Universit\`a di Milano and\\
INFN, Sezione di Milano, via Celoria 16,
I-20133 Milano, Italy }
\end{center}
\vfill

\begin{center}
{\bf ABSTRACT}
\end{center}
\begin{quote}
This is a short introduction to the study of compactifications 
of F--theory on elliptic Calabi--Yau threefolds near colliding 
singularities. In particular we consider the case of nontransversal 
intersections of the singular fibers.
\end{quote}
\end{titlepage}

\sect{Introduction}
We consider F-theory compactified to six dimensions on elliptic
Calabi--Yau threefolds and study the case in which the loci of the singular
fibers intersect each other non transversally. We show how
these singularities can be resolved by suitable blow--ups of the base
of the elliptic Calabi--Yau and explore new branches of the moduli
space where the gauge group is enhanced and the charged matter content
allows cancellation of all the anomalies.
Since these new branches are characterized by the presence of
more than one tensor multiplet, they do not admit a dual
description in terms of heterotic strings.

The talk is organized as follows: we start by reviewing some basic
concepts about F-theory compactifications to eight and six
dimensions. We briefly discuss the
conjectured F-theory/heterotic duality and, in particular, we examine the
correspondence between singularities of the compactification 
manifold on the F-theory side, and gauge group enhancements on the
heterotic side. Then we turn our attention to the resolution of nontransversal
colliding singularities by means of a suitable blow--up procedure
on the base of the elliptic Calabi--Yau manifold. In this way
we go to new branches of the moduli space that do not correspond anymore 
 to a dual perturbative heterotic theory.
Finally, in these new branches, we illustrate with an
example how to determine physical properties, namely
the gauge group enhancement and the matter content.

\sect{F-theory and its compactifications}
  
F-theory is a 12--dimensional theory that, when compactified on a torus, 
gives rise to type IIB strings in 10
dimensions \cite{vafa}. It allows to interpret in a natural geometric way
the $SL(2,Z)$ symmetry of the type IIB theory:
the complex field constructed with the R-R and
NS-NS scalars (axion $\tilde\phi$ and dilaton $\phi$) of type IIB,
$\t = \tilde\phi + i e^{-\phi}$,
can be identified with the complex modulus of the torus over which
F-theory is compactified, and the $SL(2,Z)$ symmetry is then
interpreted as the modular invariance of this torus.

New compactifications of type IIB theory are obtained starting from
F-theory and then compactifying it on manifolds that are elliptic 
fibrations 
\cite{vafa,morvafa}. We consider in particular compactifications
to 8 dimensions on elliptic $K3$, and to 6 dimensions on
elliptic Calabi--Yau threefolds.

\subsection{Compactification to eight dimensions}

We start by considering a compactification over a $K3$ which admits an
elliptic fibration.
In this simple case, the equation for the elliptic fiber (torus) is
\EQ
y^2 = x^3 + f_8(z)x + g_{12}(z)
\EN
where the two functions $f$ and $g$, polynomials of degree
8 and 12 respectively, vary over the base $\IP^1$
parametrized by $z$ \cite{vafa}.

The torus is degenerate when the discriminant of the cubic vanishes
\EQ
\D = 4(f_8)^3 + 27(g_{12})^2 = 0
\label{locus}
\EN
that is, with this choice for the degrees of $f$ and $g$, on 24
points over the sphere $\IP^1$. In fact 24 singular points are necessary
in order to satisfy the Calabi--Yau condition for the $K3$
manifold \cite{morvafa}.
On these 24 points the modulus $\t$ of the torus becomes singular \cite{vafa}
near $z=0$, $\t(z)\sim\frac{1}{2\pi}\log z$. As we go around $z=0$,
$\t\rightarrow\t + 1$, i.e.
$\tilde\phi\rightarrow\tilde\phi + 1$.
This signals the presence of a magnetically charged 7--brane
in $z=0$, filling the uncompactified space--time, with unitary
magnetic charge
\EQ
Q_{magn} = \int_{S^1} d\tilde\phi = \tilde\phi(e^{2\p i}z)
-\tilde\phi(z) = 1
\EN

The corresponding 8 dimensional theory is conjectured to be dual to the
heterotic string compactified on a torus \cite{vafa}.

\subsection{Compactification to six dimensions}

Further compactifications to six dimensions can be obtained from a
one parameter family of the above 8 dimensional dual
theories, parametrized by $\IP^1$. In this way we have the following
$N=1$ dual theories in 6 dimensions \cite{morvafa}:
$\frac{F}{\mbox{elliptic}~CY}\sim\frac{\mbox{het}}{K3}$.

In this case the equation for the elliptically fibered $CY$ threefold can be
written in the form
\EQ
y^2 = x^3 + x f(z_1,z_2) + g(z_1,z_2)
\EN
where $(z_1,z_2)$ parametrize the two-dimensional base, a $\IP^1$ bundle
over $\IP^1$.
These bundles are classified by an integer $n$ and are called
Hirzebruch surfaces $F_n$ (roughly speaking $n$ is the first
Chern class of the bundle). The trivial case corresponds to
$F_0=\IP^1\times\IP^1$.

The condition (\ref{locus}) of degeneration of the torus
gives rise to a surface, contained in the base $F_n$, which can be
interpreted as the intersection of the world--volume of
magnetically charged 7--branes and the base.
In general, $\D=0$ describes objects of complex
codimension one in the base (divisors).

Let us consider now the heterotic string theory compactified on $K3$:
the Bianchi identity $dH = tr F\wedge F - tr R\wedge  R$
imposes that the total instanton number is equal to 24.
For the $E_8\times E_8$ case we can distribute the instantons
among the two $E_8$'s as $(12+n,12-n)$.
The heterotic theory with such a configuration is conjectured
to be dual to the F-theory compactified on a $CY$ with $F_n$ as base
\cite{morvafa}. The $SO(32)$ case is assumed to be dual to
the $n=4$ case.

To test these dualities it is crucial to assume that if
the heterotic string has a gauge symmetry $G$ then on the F-theory
side the elliptic fibration must have a singularity of type $G$.
The types of singularities have been classified by Kodaira.
(For a complete list of singularities see \cite{sei}.)
From the Kodaira table we can read the type of singularity of the
$CY$ manifold simply by looking at the order of zero of the polynomials 
$f$ and $g$ and of the discriminant $\D$. 

The groups obtained in this way are always simply--laced
(ADE groups).
In some cases the singularity does not correspond to these
simply laced groups but to a quotient of them.
This happens when there is a monodromy action on the singularity
which is an outer automorphism of the root lattice \cite{sei,aspgros}.
The groups so obtained are nonsimply--laced (CBFG groups).
In this case the singularity is called non--split (it is called split
when the entire group survives).

We now present an explicit example in which we check the duality between 
the F-theory
and the heterotic string with $E_8\times E_8$ as gauge group \cite{sei}.

\subsubsection{Heterotic side}

We consider the case of heterotic string theory compactified on
$K3$ with $(12+n)$ instantons in $E_8$. In general $E_8$ is
completely broken on the hypermultiplet moduli space ${\cal H}$ of 
the gauge bundle (dim ${\cal H}$ = $30 n +112$). However, if we restrict
the instantons to sit in $SU(2)\subset E_8$ we obtain a theory
with an unbroken $E_7$ as gauge group ($E_7\times SU(2)\subset E_8$ 
is a maximal subgroup).

Standard index theorems give the matter content
\bea
\mbox{neutral hypers} & = & 2n+21 \quad (\equiv\mbox{dim of
subspace of ${\cal H}$ with $E_7$ enhancement})\nonumber\\
\mbox{charged hypers} & = & \shalf (n+8) \mbox{in the 
$\underline{56}$ of $E_7$}
\label{index}
\ena
where the $\shalf$ comes from the pseudoreality of the representation
$\underline{56}$ of $E_7$.

\subsubsection{F-theory side}

From the Kodaira singularity table one can see that the moduli
of the $E_7$ enhancement are the degrees of freedom of the two polynomials
$f_{8+n}(z_2)$ and $g_{12+n}(z_2)$ in the equation of the elliptic
fibration, that is
\EQ
(9+n) + (13+n) - 1 = 2n + 21
\EN
(the -1 comes from the rescaling of $z_1$) in agreement with the
number of neutral hypermultiplets in the heterotic calculation (\ref{index}).

Where can we read the charged matter content on the F-theory side from?
The form of the discriminant near the $E_7$ locus $\{z_1=0\}$
is $\D = z_1^9 \left(4 f^{3}_{8+n} (z_2) + o(z_1)\right)$.
This tells us that there are $(n+8)$ extra zeroes, corresponding
to $(n+8)$ 7--branes intersecting the one corresponding to $E_7$
(note, however, that the types of singularity over these extra
branes do not necessarily correspond to an extra gauge group
enhancement).
It is then natural to conclude that each charged 
$\shalf$--hypermultiplet in the $\underline{56}$ is localized
at the points of collision of the divisors. This fact was
already been conjectured before the introduction of F-theory
\cite{bersava}. A purely F-theory derivation (without the use
of duality with heterotic theory) of the charged
matter content was given in \cite{katzvafa}.

\sect{Colliding singularities}

In the last section we have described how F--theory 
compactified on elliptic Calabi--Yau threefolds with an
Hirzebruch surface as base is dual to a perturbative
heterotic string theory compactified on a $K3$.
Now we are going to explore new situations on the F--theory
side that, as we will see at the end of this section, cannot
be dual to a perturbative heterotic theory. 

By generalizing what we have seen in the last example, if
two divisors corresponding to gauge groups $G$ and $G'$ meet
each other, then one expects to have a theory with gauge group
$G\times G'$ with some matter content, either neutral or charged.
However it turns out that in many cases it is not possible to
satisfy the anomaly cancellation conditions for the F-theory \cite{sadov}.
These conditions restrict severely
the nonanomalous matter content and depend crucially upon the intersection
numbers among the divisors $D$ (topological conditions)
\bea
& & \sum_{(R_a,R_b)} n_{(R_a,R_b)} Ind (R_a) Ind (R_b) =
(D_a\cdot D_b)\nonumber\\
& & Ind (Ad_a) - \sum_{R_a} Ind (R_a) n_{R_a} = 6 (K\cdot D_a)
\nonumber\\
& & y_{Ad_a} - \sum_{R_a} y_{R_a} n_{R_a} = -3(D_a\cdot D_a)
\nonumber\\
& & x_{Ad_a} - \sum_{R_a} x_{R_a} n_{R_a} =  0
\label{acc}
\ena
where $K$ is the canonical divisor of the base. The index of a
representation $R_a$ is defined by $tr(T_a^i T_a^j) = Ind(R_a)
\d^{ij}$ and the coefficients $x$ and $y$ are defined through the
decomposition $tr_{R_a} F^4 = x_{R_a} tr F^4 + y_{R_a} (tr F^2)^2$
(`$tr$' is a trace in a preferred representation, usually the 
fundamental).
Anomaly cancellation imposes also the relation $n_H - n_V = 273 - 29 n_T$ 
among the number of vectors, tensors and hypermultiplets.

As mentioned above, in many cases the (\ref{acc}) cannot be satisfied.
For example, if $D\cdot D' = 1$ (transversal collision), the
case $SO(n)\times SO(m) (n, m\geq 7)$ cannot satisfy the first
of them (all $Ind\geq 2$). In these cases one can resolve the
singularity and satisfy the (\ref{acc}) by making a blow up of
the base \cite{due}: the intersection point is replaced by a
whole $\IP^1$ (called the exceptional divisor $E$). After the
blow--up, the two divisors corresponding to the gauge groups
do not intersect each other anymore and one can hope that the
(\ref{acc}) can be satisfied
\EQ
D\cdot D' = 1 \quad\buildrel{\mbox{blow--up}}\over\rightarrow
\quad\matrix{\hat{D}\cdot\hat{D}' & = & 0 \cr
             \hat{D}\cdot E & = & 1 \cr
             \hat{D}'\cdot E & = & 1 \cr}   
\EN
In many cases E itself becomes a component of the discriminant
locus corresponding to a new gauge group $H$.

After the blow--up one has to check that the blown--up surface
still satisfies the Calabi--Yau condition. 
This leads to the condition
\EQ
a(D) + a(D') - a(E) = 1
\label{CY}
\EN
where the coefficients $a$ depend on the singularity type
on the corresponding divisor.

So, in order to determine the new gauge group $H$, one looks for a 
coefficient $a(E)$ in Table 1
\begin{table}[ht]
\caption{Coefficients appearing in the Calabi--Yau condition}
\label{coefficients}
\begin{center}
\begin{tabular}{|c|c|c|c|c|c|c|c|c|}
\hline
none & $I_n$ & $II$ & $III$ & $IV$ & $I_n^*$ & $IV^*$ & $III^*$ &
$II^*$ \\
\hline
0 & $\frac{n}{12}$ & $\frac{1}{6}$ & $\frac{1}{4}$ & 
$\frac{1}{3}$ & $\shalf + \frac{n}{12}$ & $\frac{2}{3}$ &
$\frac{3}{4}$ & $\frac{5}{6}$ \\
\hline 
\end{tabular}
\end{center}
\end{table}
such that the (\ref{CY}) is satisfied. If $a(E)\neq  0$ then $E$ is a
component of the discriminant locus and after the blow--up the
gauge group is $G\times H \times G'$.

Now either the (\ref{acc}) can be satisfied with some matter content, 
or else other blow--ups might be needed.

The degree of freedom corresponding to the radius of the 
blown--up sphere is para\-me\-tri\-zed by the scalar of a new tensor
multiplet. Indeed, the number of tensor multiplets is given by
the number of K\"ahler classes of the base, $h^{1,1} (B)$, as
\cite{vafa,morvafa}
\EQ
n_T = h^{1,1} (B) - 1
\EN
(note that $h^{1,1} = 2$ for Hirzebruch surfaces) and the
blow--up introduces one more K\"ahler modulus. We are then in
a Coulomb branch of the moduli space, and the theory has more
than one tensor multiplet.
Note that the heterotic string has $n_T = 1$ and so after the phase
transition our theory does not have a dual  heterotic description.
The gauge group enhancement is of non--perturbative kind.

The case of transversal collisions has been studied in \cite{due}.
Let us now look at the non--transversal situation.

\sect{The non--transversal case}

For non--transversal collisions ($D\cdot D'>1$) we need to perform
more than one blow--up to resolve the singularity, in such a way that at 
the end the two divisors do not intersect each other anymore. Since
every blow--up introduces one new tensor multiplet, the phase 
transitions we are now going to explore are characterized by
$\d n_T > 1$.

For example, let us consider the case $D\cdot D' = 2$.
After a first blow--up $D$ and $D'$ intersect again each other
($\hat{D}\cdot\hat{D}' = \hat{D}\cdot E_1 = 
\hat{D}'\cdot E_1 = 1$).
Only after a second blow--up, with the introduction of a second
exceptional divisor $E_2$, the two divisors do not intersect
\EQ
\hat{\hat{D}}\cdot\hat{\hat{D}}' = \hat{\hat{D}}\cdot
\hat{E_1} = \hat{\hat{D}}'\cdot\hat{E_1} =0 \qquad
\hat{\hat{D}}\cdot E_2 = \hat{\hat{D}}'\cdot E_2 =
\hat{E_1}\cdot E_2 = 1
\EN
More generally, if $D\cdot D' = p$ we need to perform $p$ blow--ups,
with the introduction of $p$ exceptional divisors.

There are now $p$ Calabi--Yau conditions, one for every blow--up.
For example, in the $p=2$ case they are
\EQ
\left\{\matrix{a(D) + a(D') - a(E_1) & = & 1 \cr
               a(D) + a(D') + a(E_1) - a(E_2) & = & 1 \cr} 
\right.
\label{CY2}
\EN
Again, if $E_1$ and $E_2$  belong to the discriminant locus,
then we obtain the non--perturbative gauge group enhancement
to $G\times H_1\times H_2\times G'$ (with $p$ $H$ factors if 
$D\cdot D' = p$).

As an example we may consider the case of the collision between
the fibers $I_n^*$ and $I_m^*$, corresponding to the gauge groups
$SO(2n+8)\times SO(2m+8)$, with $D\cdot D' = 2$.
It is easy to see that the (\ref{acc}) cannot be verified. As described
above, we need to perform at least two 
blow--ups to resolve the singularity. Then we use (\ref{CY2})
to find the gauge group enhancement and (\ref{acc}) to
determine the matter content.

\subsection{Determining the gauge group enhancement}

From Table 1 we know that
$a(D) = \shalf + \frac{n}{12}$, $a(D') = \shalf + \frac{m}{12}$
and then, to solve the (\ref{CY2}), we have to take
\EQ
a(E_1) = \frac{n+m}{12} \qquad a(E_2) = \frac{n+m}{6}
\EN
corresponding to the fibers $I_{n+m}$ and $I_{2(n+m)}$ respectively.
Note that the singular fibers $I_{n+m}$ correspond either to
$SU(n+m)$, in the split case, or to $Sp\left(\frac{n+m}{2}\right)$,
in the non split case.
The gauge group enhancement is then
\EQ
SO(2n+8)\times\matrix{SU(n+m)\cr Sp(\frac{n+m}{2})\cr}\times
\matrix{SU(2(n+m))\cr Sp(n+m)\cr}\times SO(2m+8)
\label{groupsnm}
\EN
where the upper choices correspond to the split cases and the lower choices 
to the non split ones.

\subsection{Determining the matter content}

Let us study for simplicity the case $n=m=1$, corresponding to the enhancement
\EQ
SO(10)\times SO(10)\rightarrow SO(10)\times SU(2)\times
\matrix{SU(4)\cr Sp(2)\cr}\times SO(10)
\EN
We determine the matter content by solving (\ref{acc}). The
first of them gives the content of matter in mixed representations
of the gauge group. Note in particular that $\hat{\hat{D}}\cdot
\hat{\hat{D}}' = \hat{\hat{D}}\cdot\hat{E_1} = \hat{\hat{D}}'
\cdot\hat{E_1} = 0$ and so there is no matter in mixed representations
of the corresponding pairs of groups. We have instead
$\hat{\hat{D}}\cdot E_2 = 1$,
that means that there is matter in the mixed representation of
the first $SO(10)$ and of $SU(4)$ (or $Sp(2)$). We make a
minimal choice for the possible representations of these gauge
groups: the fundamental, the adjoint, the spinorial and the
antisymmetric.

By imposing the first of (\ref{acc}) we fix the gauge group
to be $SO(10)\times SU(2)\times Sp(2)\times SO(10)$
(picking out the non--split case) and we determine the matter
content $\shalf(10,1,4,1)$.
\\
Analogously, for the other intersections we find
$\shalf(1,1,4,10)$ and $2\times\shalf(1,2,4,1)$ or $\shalf(1,2,5,1)$.

The other equations in (\ref{acc}) give conditions that allow
us to determine the matter content in pure representations of 
the gauge groups. Note that for the two divisors
$\hat{\hat{D}}$ and $\hat{\hat{D}}'$, the right hand sides
of the second and of the third equation in (\ref{acc}) depend on the
particular choice of the base of the compactification manifold
and on the choice of the two divisors. Thus the matter content in
pure representations of $G$ and $G'$ is not universal, while
the mixed--matter content is universal depending only on the 
local structure of the colliding singularities.

Looking for a realization of $D$ and 
$D'$, one can make again a minimal choice by
imposing that there are no hypermultiplets in the adjoint.
This requires \cite{sadov} that the genus of the divisors is
zero and that
\EQ
K\cdot D = - D\cdot D -2
\EN
Very often the most convenient choice for the base of the 
elliptic $CY$ is the Hirzebruch surface $F_n$.

We then find a universal matter content
\EQ
\shalf (10,1,4,1) + \shalf (1,1,4,10) + 2\times\shalf (1,2,4,1)
\EN
and a matter part depending on our particular choice for
$D$ and $D'$
\EQ
(n_1+2)\left[(10,1,1,1) + (16,1,1,1)\right] + 
(n_2+2)\left[(1,1,1,10) + (1,1,1,16)\right]
\EN
where $n_1=D\cdot D$ and $n_2=D'\cdot D'$.

\sect{Conclusions}

We have shown with an example how non--transversal 
colliding singularities can be resolved for the $SO\times SO$ case by using
a blow--up procedure on the base of the fibration.
The analysis for general $G\times G'$ and general intersection
numbers is in progress.

We have seen that the new branches of the moduli space found in
the resolution of non--transversal colliding singularities cannot
correspond to a perturbative heterotic description. Their interpretation,
maybe in terms of instantons shrinking to zero size \cite{smallinst},
certainly deserves further study.
\vskip 0.5cm
\noindent
{\bf Acknowledgements.}
We thank M. Bershadsky and A. Johansen for useful conversations
and suggestions.
\\
This work has been supported by the European Commission TMR programme 
ERBFMRX-CT96-0045, in which the authors are associated to Torino.

\end{document}